\documentclass[rnote,traditabstract]{aa} % for the letters 
\usepackage{subfigure}                                   % (traditional abstract) 
\usepackage{epsfig}
\usepackage{natbib, graphicx}
\usepackage{amssymb, amsmath}
\usepackage{lscape}
\usepackage{wasysym}

\usepackage{amsmath}
\usepackage{color}

\usepackage{graphicx}
\usepackage{txfonts}
\begin{document}
\title{Exploring the $\alpha$-enhancement of metal-poor planet-hosting stars.
The \textit{Kepler} and HARPS samples\thanks{Table with chemical abundances is only 
available at the CDS via anonymous ftp to \texttt{cdsarc.u-strasbg.fr (130.79.128.5)} or via
\texttt{http://cdsarc.u-strasbg.fr/viz-bin/qcat?J/A+A/547/A36}}}

\titlerunning{Exploring the $\alpha$-enhancement of metal-poor planet-hosting stars}

%\thanks{}

\author{V.~Zh.~Adibekyan\inst{1} 
\and E.~Delgado~Mena\inst{1} 
%\and A.~Authors\inst{1,2,3}}
\and S.~G.~Sousa\inst{1,2}
\and N.~C.~Santos\inst{1,3} 
\and  G.~Israelian\inst{2,4}
\and \\ J.~I.~Gonz\'{a}lez Hern\'{a}ndez\inst{2,4}
\and M.~Mayor\inst{5}
\and  A.~A.~Hakobyan\inst{6,7,8}}

\institute{Centro de Astrof\'{\i}ísica da Universidade do Porto, Rua das Estrelas,
4150-762 Porto, Portugal\\
\email{Vardan.Adibekyan@astro.up.pt}
\and Instituto de Astrof\'{\i}sica de Canarias, 38200 La Laguna, Tenerife, Spain 
\and Departamento de F\'{\i}ísica e Astronomia, Faculdade de Ci\^{e}ncias da Universidade do Porto, Portugal
\and Departamento de Astrof{\'\i}sica, Universidad de La Laguna, 38206 La Laguna, Tenerife, Spain
\and Observatoire de Gen\`{e}ve, Universit\'{e} de Gen\`{e}ve, 51 Ch. des Mailletes, 1290 Sauverny, Switzerland
\and Byurakan Astrophysical Observatory, 0213 Byurakan, Aragatsotn province, Armenia
\and Department of General Physics and Astrophysics, Yerevan State University, 1 Alex Manoogian, 0025 Yerevan, Armenia
\and Isaac Newton Institute of Chile, Armenian Branch, 0213 Byurakan, Aragatsotn province, Armenia}

   \date{Received 4 August 2012 / Accepted 27 September 2012}

% \abstract{}{}{}{}{} 
% 5 {} token are mandatory
 
 \abstract
{Recent studies have shown that at low metallicities Doppler-detected planet-hosting stars tend to have high $\alpha$-content
and to belong to the thick disk. We used the reconnaissance spectra of 87 \textit{Kepler} planet candidates and data available from the
HARPS planet search survey to explore this phenomenon. Using the traditional spectroscopic abundance analysis methods, we 
derived Ti, Ca, and Cr abundances for the \textit{Kepler} stars. In the metallicity region --0.65 $<$ [Fe/H] $<$ --0.3 dex, the fraction of 
Ti-enhanced thick-disk HARPS  planet harboring stars is 12.3$\pm$4.1\%, and for their thin-disk counterparts this fraction is 
2.2$\pm$1.3\%. Binomial statistics give a probability of 0.008 that this could have occurred by chance.
Combining the two samples (HARPS and \textit{Kepler}) reinforces the significance of this result ($\rm{P} \sim 99.97 \%$).
Since most of these stars harbor small sized or low-mass planets we can assume that, although terrestrial planets can be 
found in a low-iron regime, they are mostly enhanced by $\alpha$-elements. This implies 
that early formation of rocky planets could start in the Galactic thick disk, where the chemical conditions for 
their formation are  more favorable.}{}


\maketitle
%
%________________________________________________________________________________________________________________________________

\section{Introduction}

Since the first discovery of an exoplanet around a solar-like star \citep{Mayor-95}, more than 750 exoplanets have been
discovered and more than 2300 planet candidates have been announced \citep{Batalha-12}.
Despite the rather large number, we still have to be content with a very small number of low-metallicity planet-hosting stars 
(PHSs) because of the well-established metal-rich nature of the PHSs, especially giant PHSs 
\citep{Gonzalez-98, Gonzalez-01, Santos-01, Santos-04, Fischer-05, Neves-09, Johnson-10, Sousa-11, Adibekyan-12b}.
Study of the chemical properties of metal-poor (iron-poor) PHSs is very important for understanding if there is any chemical composition
requirements  for their formation.

Recently, some studies have shown that iron-poor PHSs tend to be enhanced by $\alpha$-elements \citep[e.g.][]
{Haywood-08, Haywood-09, Kang-11, Adibekyan-12a}. This enhancement suggests that in a metal-poor regime most of the PHSs belong to the
Galactic thick disk \citep {Haywood-08, Haywood-09, Adibekyan-12a}. However, further interpretations of this result have been ambiguous.
\citet{Haywood-08, Haywood-09} suggests that metallicity is not the causative parameter that determines the presence 
of giant planets around Sun-like stars, and there might be another parameter linked to galactocentric radius 
(such as density of H$_2$). Alternatively, \citet{Adibekyan-12a} proposes that a certain chemical composition, 
and not the Galactic place birth of the stars, is the determinating factor for iron-poor PHSs to lie in the high-$\alpha$/thick-disk region
\citep[see also discussion in][]{Gonzalez-09}.

%--------------------------Fig  1-----------------------------------------

\begin{figure*}
\begin{center}$
\begin{array}{ccc}
\includegraphics[angle=270,width=0.33\linewidth]{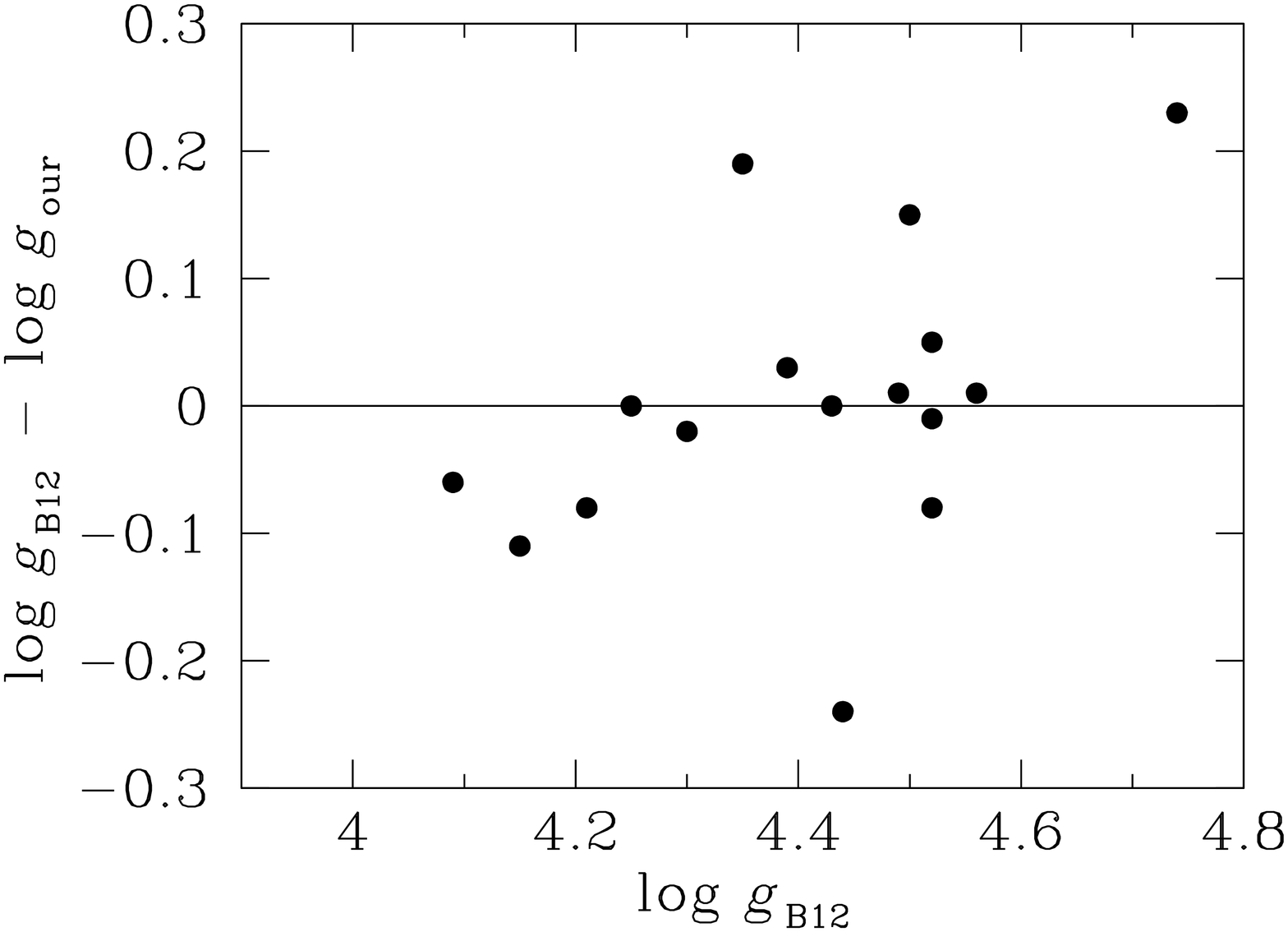}&
\includegraphics[angle=270,width=0.33\linewidth]{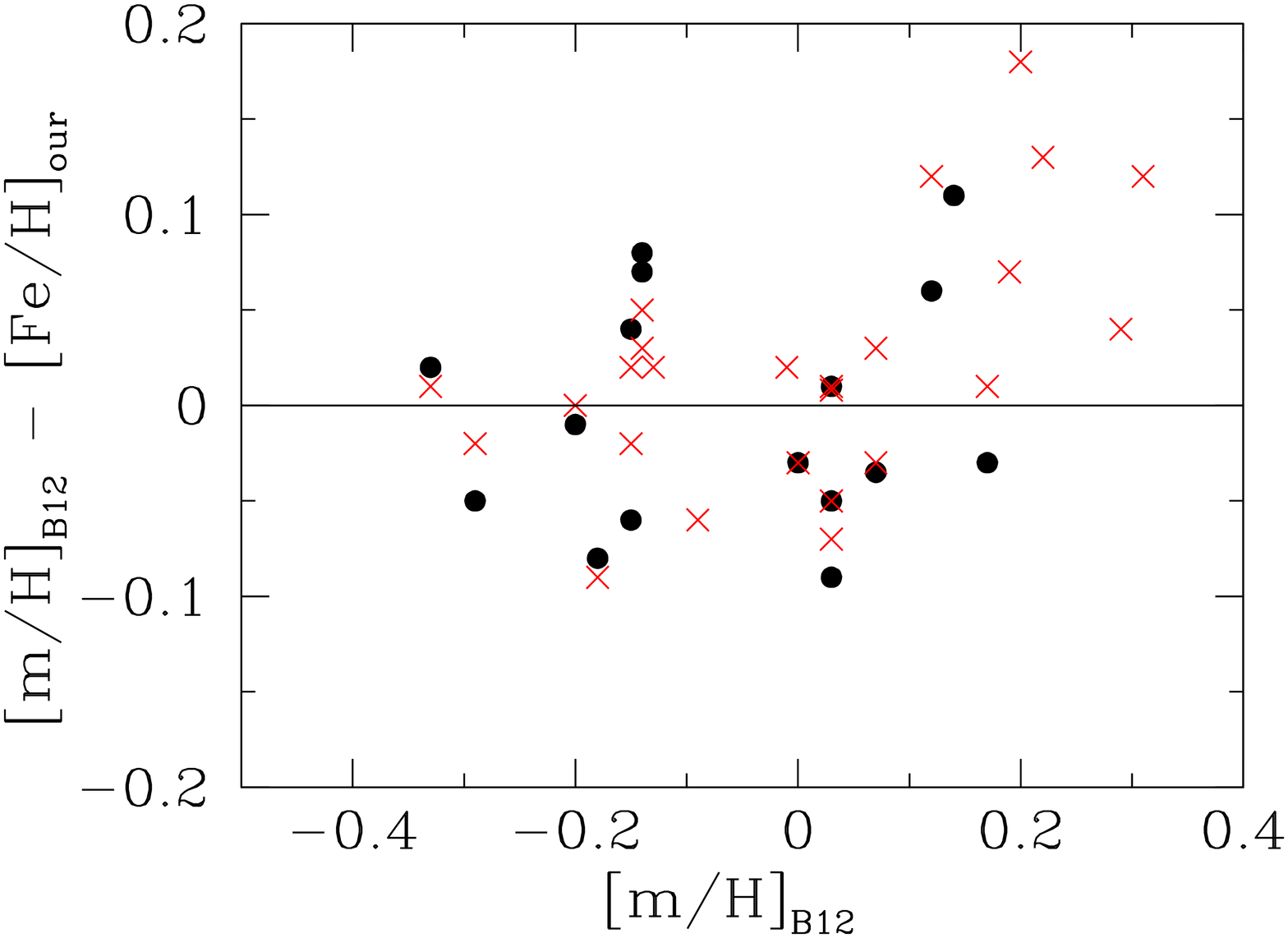}&
\includegraphics[angle=270,width=0.33\linewidth]{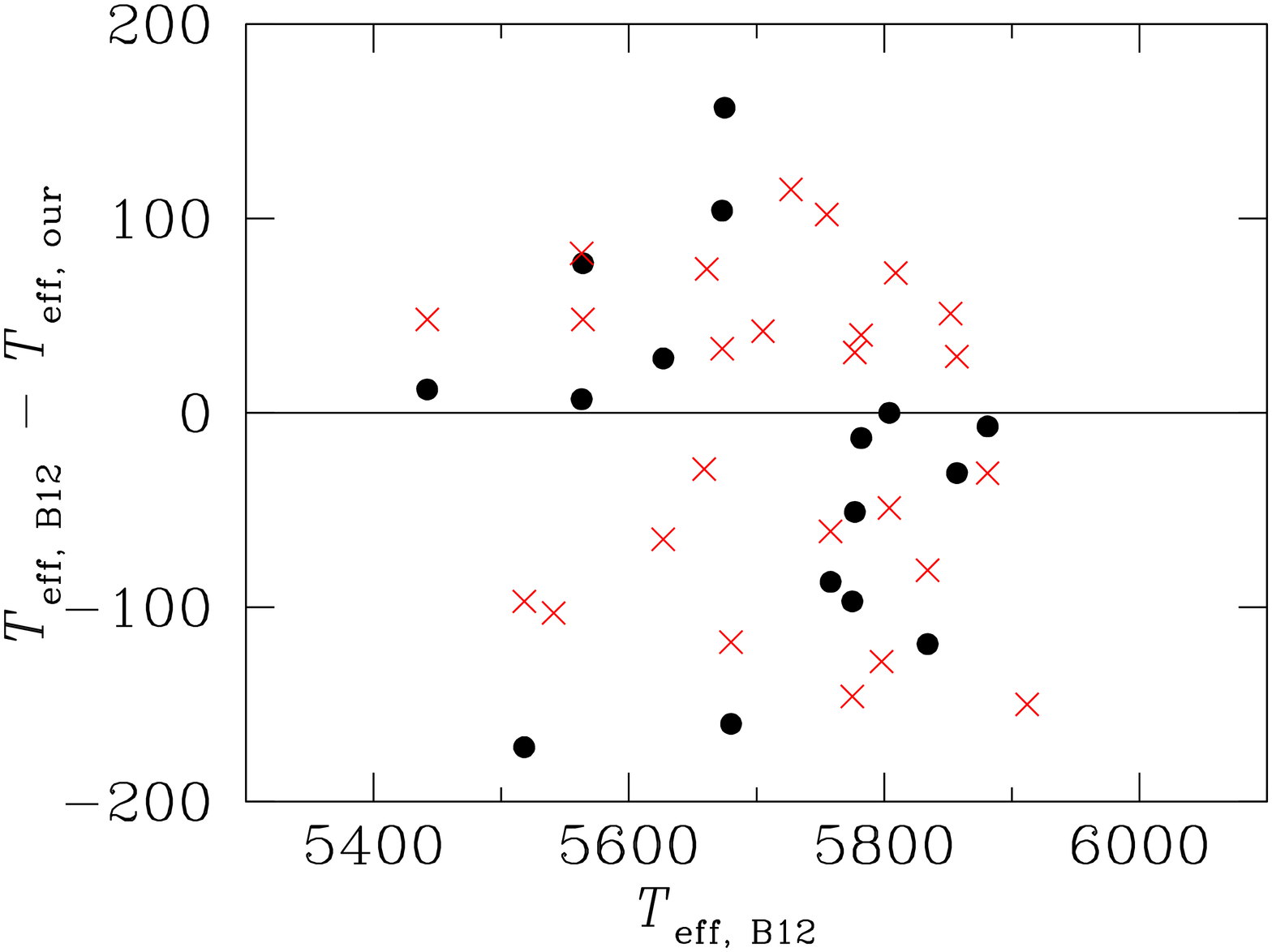}
\end{array}$
\end{center}
\caption{Comparison of the stellar parameters from B12 to those derived in this work by traditional spectroscopic analysis technique
(\emph{black dots}) and with the TMCalc code (\emph{red crosses}).}
\label{fig-1}
\end{figure*}
%-------------------------------------- -----------------------------------------

The iron content is used as a proxy for the overall metallicity of planet hosts in most studies. However, iron is not the only abundant refractory 
element in the solar system. There are other fairly abundant elements (e.g. Mg and Si) with condensation
temperatures comparable to iron \citep{Lodders-03} that are very important contributors to the composition
of dust in planet-forming regions \citep[e.g.][]{Gonzalez-09}. The $\alpha$-enhancement of iron-poor PHSs is not
unexpected, since a high [X/Fe] ratio means higher ``total metallicity'' and is moreover supported by the
theoretical studies using the core-accretion model \citep[e.g.][] {Ida-04, Mordasini-12}.
Interestingly, \citet{Adibekyan-12a} show that not only stars hosting giant planets, but also Neptune- and super-Earth class
planet hosts are also enhanced by  $\alpha$-elements in the metal-poor regime. This indicates that despite the observed
correlation between stellar metallicity ([Fe/H]) and presence of low-mass planets is very weak \citep[e.g.][] {Udry-06,Sousa-08, 
Ghezzi-10, Mayor-11, Sousa-11, Buchhave-12, Adibekyan-12b}, there is a special chemical composition required 
for their formation.

\citet[hereafter B12]{Buchhave-12} report spectroscopic metallicities of 152 stars harboring 226 exoplanet candidates
discovered by the \textit{Kepler} mission \citep{Borucki-10}, including 175 objects that are smaller than 4~$R_{\oplus}$. With precise 
stellar parameters they also provide the unprocessed extracted spectra (only the wavelength region rangs
roughly 260 $\AA$) used in their analysis. In this paper we have
used these spectra and data from \citet{Adibekyan-12b} to explore the possible chemical anomalies (i.e., $\alpha$-enhancement) described above. 
%The letter is organized as follows: Sect.~\ref{sample} introduces the sample and stellar parameters. Sect.~\ref{refractory} 
%contains the abundance analysis. The discussion of our results and main conclusions are presented in  
%Sect.~\ref{discussion}.

\section{The sample and atmospheric parameters}
\label{sample}

Recently, B12  used the reconnaissance spectra obtained by the \textit{Kepler} Follow-up Observing Program to 
derive the stellar parameters for 152 stars (\textit{Kepler} Objects of Interest - KOI) hosting 226 planet candidates. 
These stars were chosen out of about 560 KOIs (having reconnaissance spectra) as not very cool, 
$\emph{$T{}_{\mathrm{eff}}$} > 4500$  K, and not fast rotators $\upsilon$~sin$i$ $<$ 40 km s$^{-1}$, which 
allowed them to get more robust results. 
The spectra used in B12 were obtained with high-resolution spectrographs on medium-class telescopes, and
some highest priority planet candidates (those of nearly Earth-sized) were observed with the HIRES spectrograph \citep{Vogt-94}
on the 10-m Keck I telescope at Mauna Kea, Hawaii. The spectra have low/modest signal-to-noise ratio (S/N; S/N per
pixel $>$ 15). We refer the reader to B12 for more details.

Our main goal is to study the chemical abundances of individual elements of planet-hosting dwarf stars with 
standard spectroscopic analysis techniques using equivalent widths (EW) of the spectral lines. This technique
requires higher S/N and lower $\upsilon$~sin$i$ than used in B12, where the stellar parameters were determined
by matching an observed spectrum to a library grid of synthetic model spectra. For this reason we first selected 
only stars with $\upsilon$~sin$i$ $<$ 10 km s$^{-1}$ and $\log{g}$ $>$ 4.0 dex, which reduced the sample to 117 KOIs.
Because many of these objects have multiple observations with different instruments, we were not limited in S/N
(there is no numerical criterion), but during determination of chemical abundances we selected only those with 
reliable EW measurements (see Sect.~\ref{refractory}). Our final sample, for which we were able to determine elemental 
abundances, consists of 87 PHSs with 146 planet candidates.

\subsection{Testing the stellar parameters}

As mentioned before, in B12 the stellar parameters (\emph{$T{}_{\mathrm{eff}}$}, $\log{g}$, [m/H], and $\upsilon$~sin$i$) 
were derived by cross-correlating the observed spectra against synthetic model spectra. To test how these parameters
match those obtained with traditional techniques using EWs, we selected the stars with highest S/N observed with Keck~I
(41 stars). For these stars we applied two different techniques to derive the stellar atmospheric parameters.

In the first approach the spectroscopic stellar parameters and metallicities were derived based on the EWs 
of the  \ion{Fe}{i} and  \ion{Fe}{ii} weak lines by imposing excitation and ionization equilibrium \citep[e.g.][] {Sousa-11}.
The spectroscopic analysis was completed assuming local thermodynamic equilibrium (LTE) with a grid of Kurucz atmosphere 
models \citep{Kurucz-93}, and making use of a recent version of the MOOG%
\footnote{ The source code of MOOG2010 can be downloaded at \texttt{http://www.as.utexas.edu/$\sim$chris/moog.html}%
} radiative transfer code  \citep{Sneden-73}. The quality of these spectra was high enough to measure the EW using the ARES%
\footnote{The ARES code can be downloaded at \texttt{http://www.astro.up.pt/sousasag/ares}%
} code \citep{Sousa-07}.
Unfortunately, in the used spectral window there are 25 \ion{Fe}{i} and 5 \ion{Fe}{ii} lines at most, which is not always enough to 
get precise parameters. 

In the second approach, we used the TMCalc code \citep{Sousa-12}. This code uses a recent line-ratio calibration 
\citep{Sousa-10} to estimate the \emph{$T{}_{\mathrm{eff}}$}. Then, it uses a new direct spectroscopic calibration based 
on weak \ion{Fe}{i} lines, which are expected to be less dependent on surface gravity and microturbulence to estimate [Fe/H]. 
The code was combined with ARES to estimate both the spectroscopic stellar \emph{$T{}_{\mathrm{eff}}$} and [Fe/H]. 
With this code we determined the mentioned parameters for 28 KOIs.

%--------------------------------Fig 2 -----------------------------------------------------------------------
\begin{figure}
\centering
\includegraphics[width=1\linewidth]{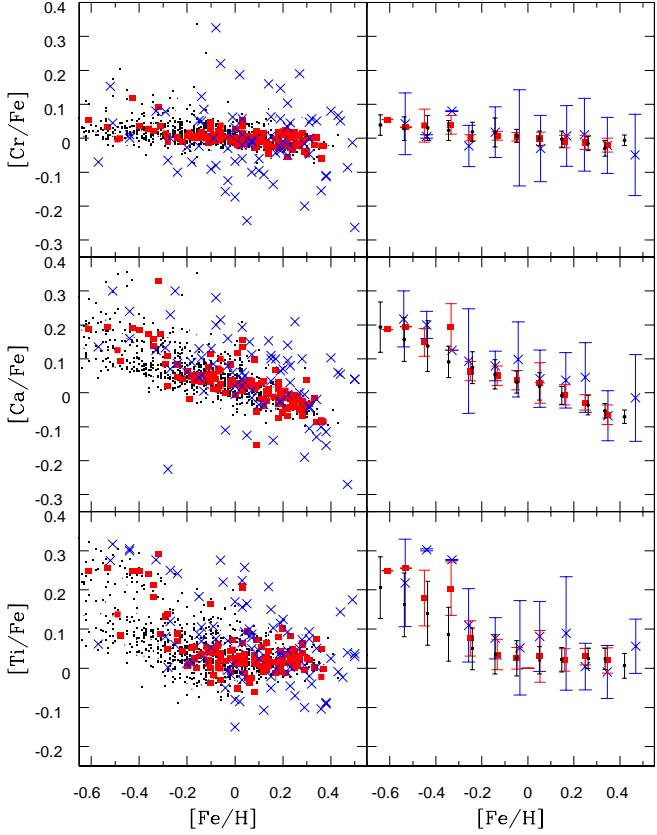}\caption
{[X/Fe] abundance ratios against [Fe/H] for the stars with and without planets from HARPS and \textit{Kepler} surveys. 
The symbols and error bars (right panel) indicate the average and standard deviation, respectively, of each bin (0.05 dex). 
The \emph{red squares} and \emph{blue crosses} represent stars with planets detected by HARPS and \textit{Kepler}, respectively. 
The \emph{black dots} refer to stars without a planetary companion.}
\label{fig-2}
\end{figure}
%-------------------------------------------------------------------------------------------------------

Comparing the stellar parameters derived with the two methods we found strong discrepancies for the majority of the stars.
This is not surprising because the first method is designed to use a long list of \ion{Fe}{i} and  \ion{Fe}{ii} lines. However, we found
good agreement between the [Fe/H]%
\footnote{The metallicity in B12 is denoted as [m/H] and represents a mix of metals assumed to be the same as 
the relative pattern of the abundances in the Sun.%
} and \emph{$T{}_{\mathrm{eff}}$} determined with the TMCalc code and B12, except for three
stars for which the differences in \emph{$T{}_{\mathrm{eff}}$} are higher than 250 K. The comparison
of these two parameters (except the three mentioned stars) is shown in Fig.~\ref{fig-1}. 
As can be seen, the two results agree well, although it seems there is a small systematic trend at higher metallicities 
\citep[see also][]{Torres-12}.
The standard deviations of the 
differences of \emph{$T{}_{\mathrm{eff}}$} and [Fe/H] between the two estimations are 
$\sigma$(\emph{$T{}_{\mathrm{eff}}$}) = 82 K and  $\sigma$([Fe/H]) = 0.07 dex (25 stars).
This independent analysis also shows that the TMCalc code works well
and can be very useful for determining precise parameters for large amounts of data.

Assuming that the most reliable parameters were obtained for the stars for which our two methods gave similar results 
($\mid$$\Delta$\emph{$T{}_{\mathrm{eff}}$}$\mid$ $<$ 100 K and $\mid$$\Delta$[Fe/H]$\mid$ $<$ 0.1 dex), we compared our atmospheric parameters 
(obtained with the first method) with those obtained in B12. The comparison for these 16 PHSs is presented in Fig.~\ref{fig-1}. 
As can be seen, the two results agree well. The standard deviations of the differences of the stellar 
parameters derived in B12 and with our first method are $\sigma$(\emph{$T{}_{\mathrm{eff}}$}) = 100~K,   
$\sigma$([Fe/H]) = 0.09 dex, and $\sigma$($\log{g}$) = 0.16 dex (16 stars). Based on this consistency we use the atmospheric 
parameters from B12 in our analysis here. 

\subsection{Microturbulence}

The traditional spectroscopic abundance analysis methods using the MOOG code and a grid of Kurucz model atmospheres 
requires four stellar parameters: [Fe/H], \emph{$T{}_{\mathrm{eff}}$}, $\log{g}$, and microturbulence velocity
(\emph{$\xi{}_{\mathrm{t}}$}). B12 provides only the first three atmospheric parameters.
Earlier studies of FGK dwarfs have shown that \emph{$\xi{}_{\mathrm{t}}$} depends on \emph{$T{}_{\mathrm{eff}}$} and $\log{g}$
\citep[e.g.][] {Nissen-81, Reddy-03, Prieto-04}. Using the spectroscopic \emph{$\xi{}_{\mathrm{t}}$}
of 1111 FGK dwarf stars from the high-S/N and high-resolution HARPS data  \citep{Sousa-08,Sousa-11,Sousa-11a}, 
we derive the following expression 

\begin{eqnarray}\nonumber
\xi_{t} & = & 9.185 - 3.161\times10^{-3}T_{eff} +  3.436\times10^{-7}T_{eff}^{2} \left. \right.\\
&& \left. - 0.316\times\log{g}  \,\ , \right.
\end{eqnarray}
expecting an rms error of 0.18 km s$^{-1}$. Here the \emph{$\xi{}_{\mathrm{t}}$} is in km s$^{-1}$, and 
\emph{$T{}_{\mathrm{eff}}$} and $\log{g}$ are in their traditional units. The stars in the sample have \emph{$T{}_{\mathrm{eff}}$} 
ranging from 4500 to 6500 K, $\log\,g$ ranging from 3 to 5 dex, and [Fe/H] ranging from --1.4 to 0.5 dex.
We use this expression to calculate the \emph{$\xi{}_{\mathrm{t}}$} for our \textit{Kepler} sample stars.
%-------------------------------------- -----------------------------------------

\section{Refractory elements in planet-hosting stars}
\label{refractory}

The reconnaissance spectra used in this work cover from about 70 to 260 $\AA$ around $\lambda \approx 5200 \, \AA$
(see B12 for more details). In this spectral range we have four  \ion{Ti}{i} and one \ion{Ti}{ii} line, 
five \ion{Cr}{i} lines, and two \ion{Ca}{i} lines at most 
\citep[for more details about the lines see][]{Neves-09}. For these lines the EWs were measured using a
Gaussian fitting procedure within the IRAF%
\footnote{IRAF is distributed by National Optical Astronomy Observatories, operated by the Association of Universities for
Research in Astronomy, Inc., under contract with the National Science Foundation, USA.}
\texttt{splot} task. After measuring the EWs of spectral lines for the three elements, 
the chemical abundances were derived  using an LTE
analysis with the Sun as reference point with the MOOG spectral synthesis code  \citep{Sneden-73} 
and a grid of Kurucz ATLAS9 plane-parallel model atmospheres \citep{Kurucz-93}.
The reference abundances used in the analysis were taken from \citet{Anders-89}.

We derived the average abundances of the stars from all the available spectra taking their quality into account. 
Additionally, we removed the lines that were very 
different from the average value. We are inclined to trust the results obtained from the spectra with higher S/N.
For some stars there was only one low-quality spectra available with unreliable abundance results (very different 
abundances from different lines). We excluded these stars from our sample. 
At the end of this procedure, we finished with 75 KOIs with Cr abundances, 75 KOIs with Ca, and 86 KOIs
with Ti abundances. The final sample consists of 87 KOIs with at least one abundance result.

In Fig.~\ref{fig-2} we show the [X/Fe] abundance trends relative to [Fe/H] for the total \textit{Kepler} sample.
For comparison, the stars hosting planets and field dwarfs without planets from the sample of \citet[][HARPS sample]{Adibekyan-12a}
are also depicted. In the same plot the averages and standard deviations of [X/Fe] ratios for each [Fe/H] bin (0.05 dex)
are also displayed (right panels). These figures obviously show that the \textit{Kepler} and HARPS PHSs
have the same behavior, although the scatter of \textit{Kepler} stars is much higher. 
We note that the methodology used for deriving the chemical abundances is exactly the same for both samples (\textit{Kepler} and HARPS),
and the errors due to the applied methodology (the errors induced by uncertainties in the model atmosphere parameters) are the same.
This means that the heavy scatter found for \textit{Kepler} 
PHSs is not likely to be astrophysical, and it indicates the low precision of the EW measurements. The average scatter 
for KOIs in the metallicity region --0.1 to 0.5 dex (in this regime the astrophysical scatter due to the Galactic 
chemical evolution is low) is about 0.11, 0.10, and 0.10 dex for Cr, Ca, and Ti, respectively.  
For comparison the same scatters for HARPS PHSs are 0.02, 0.05, and 0.04 dex for Cr, Ca, and Ti,  
respectively,  and for non-host stars these values are 0.03, 0.04, and  0.04 dex.

For the \textit{Kepler} stars the line-to-line scatter errors we estimated are about 0.09, 0.10, and 0.10 dex for [Cr/H], [Ca/H], and [Ti/H], respectively.
Taking the errors induced by uncertainties into account in the model atmosphere parameters (0.03, 0.04, and 0.04 for [Cr/H], [Ca/H], and [Ti/H])
and the errors in metallicities (maximum 0.08 dex), we can expect the error of about 0.13, 0.13, and 0.12 dex for [Cr/Fe], [Ca/Fe], and [Ti/Fe]
ratios, respectively.

%--------------------------------Fig 3 -----------------------------------------------------------------------
\begin{figure}
\centering
\includegraphics[width=1\linewidth]{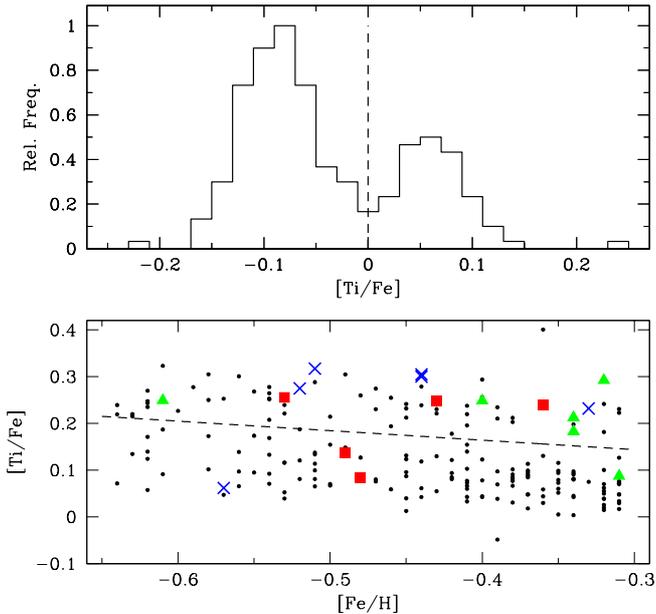}\caption
{The thin/thick separation histogram after subtracting the separation curve based on HARPS data ($top$).
[Ti/Fe] against [Fe/H] for the metal-poor stars ($bottom$). The \emph{black dashed} line indicates the separation
between the thin- and thick disks. The \emph{blue crosses} represent \textit{Kepler} planet-hosting candidates.
The \emph{red squares} refer to the Jovian hosts and \emph{green triangles} refer to the stars exclusively hosting Neptunians and super-Earths
detected by HARPS. The \emph{black dots} refer to the stars without a planetary companion.}
\label{fig-3}
\end{figure}
%-------------------------------------------------------------------------------------------------------

\section{Discussion and conclusions}
\label{discussion}

%In our previous study \citep{Adibekyan-12b} based on HARPS sample we showed that at metal-poor regime the frequency of 
%PHSs in the chemically defined thick disk is higher than in the thin disk.
As already noted, the main goal of this work is to study the behavior of \textit{Kepler} PHSs at low metallicities, i.e.,
how these planet hosts are enhanced by $\alpha$-elements (relative to iron). In this study we used Ti as a proxy of overall $\alpha$-content.
We chose not to combine Ca with Ti, because the separation between the thin and thick disks based on the [Ca/Fe] is not as
clear as for [Ti/Fe] \citep[e.g.][] {Bensby-03,Neves-09,Adibekyan-12a}.
In Fig.~\ref{fig-3} (bottom), we plot the distribution of stars both from \citet{Adibekyan-12b} and KOIs from this 
sample in the [Ti/Fe] vs. [Fe/H] space. As can be seen, the stars are clearly separated into two groups according to their
Ti content. The top panel of Fig.~\ref{fig-3} shows 
the separation histogram after subtracting the thin - thick disk separation curve. 
We note that this separation is based only on HARPS sample stars. 

As already found  \citep{Adibekyan-12b} in the metal-poor regime, the frequency of PHSs in the chemically defined 
thick disk is higher than in the thin disk \citep[for details of the chemical separation see][] {Adibekyan-11}. 
Eight stars out of 65 are hosting planets (12.3$\pm$4.1\%) above the
separation line, and only three stars out of 136 harbour planets (2.2$\pm$1.3\%) in the thin disk (low Ti-content). 
Because we have both the number of stars hosting planets and stars without detected planets from the same survey (HARPS),
we can calculate the probability that 8 (or more) out of 11 PHSs found in this metallicity range belong to the thick disk
(Ti-enhanced). The binomial statistics give a probability of 0.008 that this could have occurred by chance.

In the metallicity region --0.65 $<$ [Fe/H] $<$ --0.3 dex, there are six \textit{Kepler} PHSs, five of which lie above the [Ti/Fe] separation line
(see Fig.~\ref{fig-3}).
If we assume that the \textit{Kepler} targets also have a similar distribution in the [Ti/Fe] vs. [Fe/H] space%
\footnote{This can be very rough assumption because of the different fields of \textit{Kepler} and HARPS.},
as we observed for HARPS (32\% stars lie above and 68\% bellow the separation line), 
we can see that the probability of finding five (or more) out of six  KOIs to be Ti-enhanced is about 1.6\%. 
Moreover, if we combine both samples (13 Ti-enhanced PHSs out of 17) the binomial probability gives a value of 3 $\times$ 10$^{-4}$.
This statistically significant result (3$\sigma$) suggests that in the metal-poor regime PHSs tend to have high Ti-content.

Some of the \textit{Kepler} planetary candidates are expected to be false positives (FPs) that do not turn out to be transiting planets.
Two stars in our metal-poor \textit{Kepler} sample host three planets, one hosts four planets, one hosts five, and only two stars
host only one planet. The probability that all our planet candidates that are in the multiple systems are FPs is very low
\citep[e.g.][] {Lissauer-12}, and the FP probabilities for the two single-planet candidates are 2\% for KOI105.01 and 4\%  for KOI373.01
\citep{Morton-11}. The probability that one of the KOIs or both of them hosts FPs is thus about 6\%.
If even one of the \textit{Kepler} planet candidates is an FP, then we have four Ti-enhanced PHS out of five, for which  the binomial
statistics give a probability of about 4\% that this could have occurred by chance.
Taking the 6\% FP probability in our sample into account we can calculate a binomial probability of about 98\% that metal-poor \textit{Kepler} 
planet candidates are really (not randomly) Ti-enhanced. The same statistics give a probability of about 99.96\% ($\approx 3\sigma$) 
for the combined (\textit{Kepler} and HARPS) sample.

The enhancement of metal-poor PHSs by Ti, suggests that these stars are also enhanced by other $\alpha$-elements
\citep[e.g.][] {Neves-09,Adibekyan-12a}, including Mg and Si, which may be very important contributors to the composition
of dust in planet-forming regions and represent the principal components of rocky-type planets
\citep[e.g.][] {Gonzalez-09,Adibekyan-12b}. 

B12 suggested that small planets may be widespread in the disk of our Galaxy, with no special requirement of enhanced metallicity
for their formation. Interestingly, all six KOIs harbor small-size planets with a radius smaller than 4~$R_{\oplus}$. 
We note that six out of eleven PHSs (in the low-metallicity region) from the HARPS sample exclusively host Neptunes or super-Earth-class planets 
(less than 30 $\emph{M}_{\oplus}$), and only one lies in the Ti-poor region. The binomial statistics give a probability of about 
4 $\times$ 10$^{-4}$ that ten (or more) out of twelve small/low-mass planet hosts (combined sample) could randomly be Ti-enhanced.
Considering the possibility that \textit{Kepler} planet candidates are FPs, the  binomial statistics give a probability of about 99.95\%
($\approx 3\sigma$) that Ti-enhancement of small/low-mass planet hosts is not accidental. 
Our results suggest that, although terrestrial planets can be found in a low-iron regime, they are mostly enhanced
by other metals. Moreover, high-$\alpha$ content at low-metallicities also hints that they belong to the Galactic thick disk, 
which implies that early formation of rocky planets could have started in the Galactic thick disk,  where the chemical conditions for 
their formation are  more favorable.

%
%________________________________________________________________
\begin{acknowledgements}

{This work was supported by the European Research Council/European Community under the FP7 through Starting Grant agreement 
number 239953. N.C.S. also acknowledges the support from the Funda\c{c}\~ao para a Ci\^encia e a Tecnologia (FCT) through 
program Ci\^encia\,2007 funded by FCT/MCTES (Portugal) and POPH/FSE (EC), and in the form of grant 
PTDC/CTE-AST/098528/2008. V.Zh.A., S.G.S., and E.D.M are supported by grants SFRH/BPD/70574/2010, 
SFRH/BPD/47611/2008, and SFRH/BPD/76606/2011 from the FCT (Portugal), respectively.
J.I.G.H. and G.I. acknowledge financial support from the Spanish Ministry project MICINN AYA2011-29060 and J.I.G.H.
also from the Spanish Ministry of Science and Innovation (MICINN) under the 2009 Juan de la Cierva Program.}
\end{acknowledgements}
%________________________________________________________________

\bibliography{keplerbib}

\end{document}